\documentstyle[preprint,aps,prb,epsf,amsbsy]{revtex}
\begin{document}
\draft
\newcommand{\ve}[1]{\boldsymbol{#1}}



\title{Some Peculiarities of Proton Transport in Quasi-One-Dimensional 
Hydrogen-Bonded Chains}
\author{Natalie I. Pavlenko\\
Institute for Condensed Matter Physics\\
1, I. Svientsitsky Str., UA-79011, Lviv, Ukraine}
\date{\today}
\maketitle
\begin{abstract}
The protonic conductivity in the hydrogen bonded chains is investigated 
theoretically in the framework of the two-stage transport model. The strong 
interactions with optical phonon stretching mode are considered. We obtain 
the transition from the insulator to the metal-type state from the 
temperature dependencies of the hopping conductivity and analyze the 
influence of the specific Grotthuss mechanism on the transition character. 
We investigate also the main peculiarities in the frequency dependencies of 
the band and hopping conductivity parts which appear due to above-mentioned 
two-stage process of the proton migration along the chain.
\end{abstract}


\section{INTRODUCTION}

The investigations of transport phenomena in hydrogen-bonded materials 
attract much attention due to their fundamental role in a variety of systems 
which are important for industrial, biological and other applications. It is 
well known that electric currents in these compounds are carried out almost 
entirely by mobile protons. On the one hand, this refers to 
proteins and biomembranes with one-dimensional channels for proton 
transport (for instance, proton pathways in bacteriorhodopsin) in which the 
proton transfer is directly involved in many biochemical processes 
\cite{hadzi}. 
Such a materials are well known as 
sufficiently well protonic conductors with the conductivity in chain 
direction which exceeds 10$^3$ times the conductivity in perpendicular 
direction \cite{kawada}. On the other hand, superprotonic phases 
discovered in some H-bonded systems are characterized by drastic increase 
of the protonic conductivity up to the values about 0.1~$\Omega^{-1} 
\cdot$~cm$^{-1}$.
It is generally accepted \cite{onsager,belushkin,zetterstrom} that the 
two-stage conduction mechanism is required to sustain 
the proton transport. The 
intrabond proton tunnelling along the hydrogen bridge is connected with the 
transfer of ionic positive and negative charged defects, whereas the 
intermolecular proton transfer due to 
the reorientations of molecular group with proton leads to the breaking
of the hydrogen bond and creation of a new one between another pair of
molecular complexes (so-called orientational Bjerrum defects). 
It should be noted that the formation of the hydrogen
bridge induces the distortion of groups involved in 
hydrogen bonding towards the proton that results in the shortening of this 
bond \cite{pietraszko}. By this means the 
protonic polaron is localized between distorted ionic groups at 
sufficiently low temperatures, giving rise in 
particular to the dimerized low-temperature phases of some superprotonic 
conductors \cite{pietraszko2}. As has been shown in 
Ref.~\onlinecite{pavlenko}, the small polaron is formed due to the strong 
coupling of proton with the optical stretching vibration modes of the 
molecular groups. 

Since the transport process has a two-stage character, the problem of the 
analysis of each stage contribution to the total proton mobility is of 
particular interest as far as there exist a systems with different relative
concentration of ionic and orientational defects. 
It should be pointed out that 
whereas in systems like KH$_2$PO$_4$ a rapid dielectric relaxation is 
attributed to an abundance of mobile ionic defects and the rotation is 
sluggish, in ice the orientational D and L faults dominate the dielectric 
relaxation due to their greater numbers that limits the conductivity in ice.
Indeed, the latter fact should manifest itself in the peculiarities of 
measured conductivity, and careful theoretical investigation could provide 
a possibility to support existing conjectures concerning the transport 
mechanism. 

We start in this work with a double-minimum model for the protonic 
potential, which is modified by strong interaction with lattice vibrations.
Aside from the proton motion in a double-well potential within the bond, we 
assume the interbond proton hopping which makes possible migration 
of protonic polaron along the chain. We neglect here the interproton 
correlation effects that can be valid in the case of low proton 
concentration in chain. However, we take into account the possibility of proton exchange 
between our selected chain and surrounding.

We study in this work the influence of ionic group displacements on the 
proton subsystem behavior.
It is worthy to note that since the proton contribution into the electric 
current dominates in considered here H-bonded materials, the interaction of 
proton subsystem with the lattice distortions may lead to localization of 
protons and the system would therefore be an insulator. However, as we show 
in this work, the temperature dependence of the conductivity $\sigma$ 
changes from that typical of an insulator (decrease of $\sigma$ upon 
lowering $T$) to that typical of a conductor (increase of $\sigma$ upon 
lowering $T$). This effect points out to the existence of the so-called 
"metal"-insulator transition which was observed in the electronic 
semiconductors \cite{kravchenko}. 
Moreover, the proposed here theoretical approach enables us to 
analyze the influence of the specific transport mechanism on the 
transition character. 
We show that in our case the transition to the "metallic" state with 
temperature increase has the peculiar two-step character which attributed to 
the two-stage nature of the proton migration process. Besides the specific 
temperature dependence, the above-mentioned complex transport mechanism 
manifests itself in the frequency dependencies of $\sigma$. Although we 
neglect initially the correlations between protons, it appears from the 
dynamical conductivity behavior, that the proton motion in the chain is a 
strongly correlated process (the neighbouring protons tend to move in pairs 
along the chain).
While the first step of our analysis consists in the 
quasi-one-dimensional chains study, we believe our results can also be 
relevant for other hydrogen-bonded materials. 

\section{DESCRIPTION OF THE MODEL}

\begin{figure}[htbp]
\epsfxsize=9.cm
\epsfysize=3.cm
\centerline{\epsffile{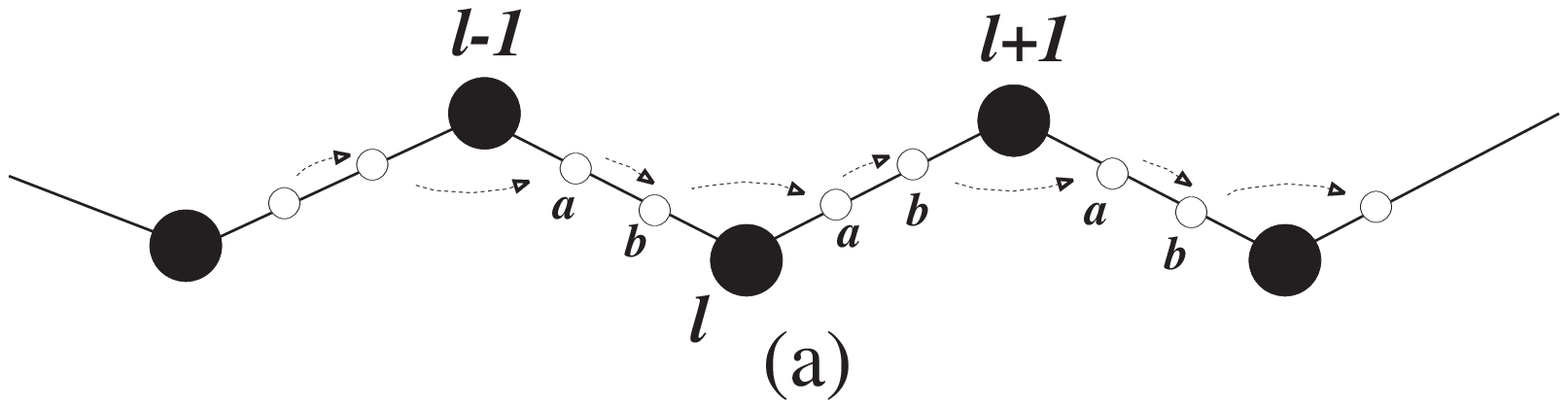}}
\null\vspace{0.1in}
\epsfxsize=7.5cm
\epsfysize=5.cm
\centerline{\epsffile{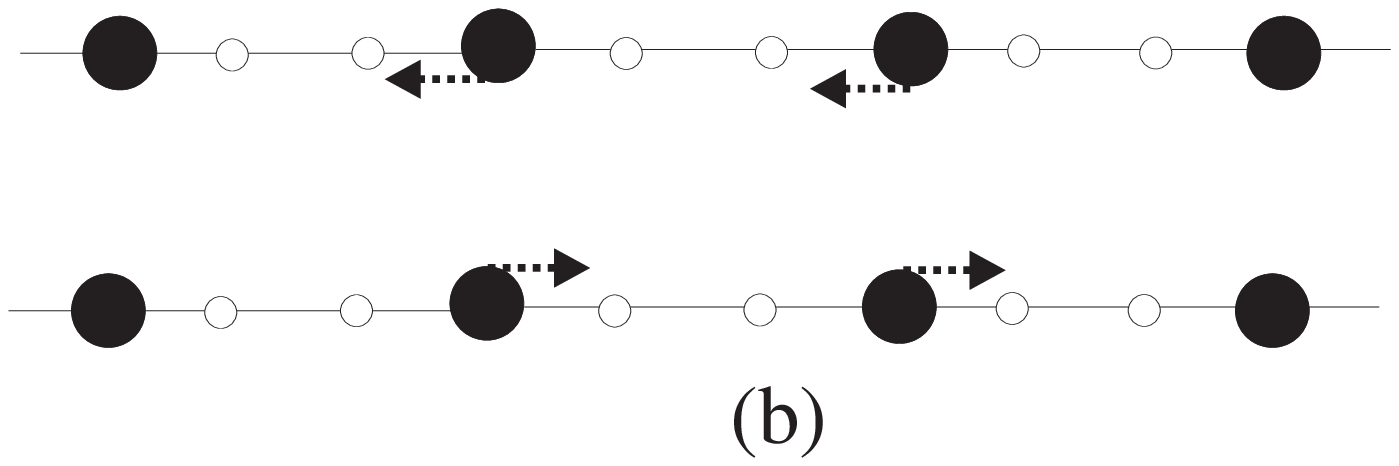}}
\caption{}
\label{fig1}
\end{figure}

The object of our consideration is the infinite chain shown in 
Fig.~\ref{fig1}(a). However, to avoid the geometric complexities introduced 
by the kinks in such a zig-zag chain, we consider in our model linear chain 
(see Fig.~\ref{fig1}(b) where two neighboring chains are shown). The process 
of the proton transfer in the double-well H-bond potential is represented as 
the quantum tunnelling between two proton states with the intrabond transfer 
integral $\Omega_T$
\begin{equation}
\Omega_T\sum_l(c_{la}^+ c_{lb}+c_{lb}^+ c_{la}), \label{h1}
\end{equation}
where $c_{l\nu}^+$, $c_{l\nu}$ denote proton creation and annihilation 
operators in the position ($l$, $\nu=a,b$) of the chain. Besides that, we 
describe the interbond reorientational proton hopping in two-level 
approximation as the quantum tunnelling effect with the hopping amplitude 
$\Omega_R$
\begin{equation}
\Omega_R\sum_l(c_{l+1,a}^+ c_{lb}+c_{lb}^+ c_{l+1,a}).\label{h2}
\end{equation}
In this way within the framework of orientational-tunnelling model proposed 
in Ref.\ \onlinecite{jps} the two-stage proton migration 
mechanism can be considered as the sequential migration of the ionic and 
orientational defects. The proton energy spectrum and dielectric properties
of the finite H-bonded complexes were analyzed on the basis of exact 
diagonalization procedure \cite{jps}.

As far as such a double-well chain is just a structural component of the 
system we also admit a possibility of proton exchange between the chain and 
surroundings by considering the system thermodynamics in the framework of 
the grand canonical ensemble with inclusion of the proton chemical potential
\begin{equation}
-\mu \sum_{l,\nu} n_{l\nu} \label{h3}
\end{equation}
which is to be determined at the given proton concentration in the chain 
from corresponding equation for the chemical potential. 

Our main interest here is to analyze the influence of the longitudinal
optical ionic group vibration modes on the proton subsystem transport
properties. We consider the optical in-phase stretching vibrations of
ionic groups in the chain which induce their displacements with respect
to the surrounding chains as identified in Fig.~\ref{fig1}(b) by dashed 
arrows. The interactions of protons with this optical mode cause the 
difference of the potential minima ($l$, $a$) and ($l$, $b$) depth within 
the bond 
\begin{equation}
\sum_{l,q} \tau_l(q)(n_{la}-n_{l-1,b})(b_{q}+
b_{-q}^+). \label{h5}
\end{equation}
Here $\tau_l(q)=g\sqrt{{\hbar}/{2MN\omega(q)}} \exp[iqld]$ where $g$ 
is the corresponding coupling constant, $M$ is the effective ionic 
group mass, $N$ denotes the number
of hydrogen bonds in chain and $d$ is the lattice 
parameter. The optical phonon branch creation and annihilation
operators are denoted by $b_{q}^+$ and $b_{q}$ respectively.
Furthermore we assume the harmonic approximation for the 
lattice vibration energy
\begin{equation}
\sum_{q} \hbar\omega(q) b_{q}^+ b_{q}. \label{h6}
\end{equation}

First of all let us consider the case of the isolated chain without 
a coupling to the phonon bath. Since the Hamiltonian (\ref{h1})-(\ref{h3}) 
can be exactly diagonalized, the proton energy spectrum 
\begin{equation}
\varepsilon_\nu (k)=\pm |t_{k}|, \hspace{0.05in}
t_{k}=\Omega_T+\Omega_R {\rm e}^{-ikd}
\end{equation}
forms two energy bands with the bandwidth 
$\Delta \varepsilon=\Omega_T+\Omega_R-|\Omega_T-\Omega_R|$. The energy gap
in this case is $\Delta_{ab}=2|\Omega_T-\Omega_R|$. Eliminating one of the 
elementary transport process by setting the hopping amplitude $\Omega_T=0$
or $\Omega_R=0$, we can see that both of the energy bands degenerate into 
the two energy levels and the quantum fluctuations between these two system 
states could be derived. It is clear that in the case when 
$\bar{n}=\frac{1}{N} \sum\limits_{l\nu} \langle n_{l\nu} \rangle=1$ 
(one proton is averaged within 
the bond) the lower band is filled and the chemical potential $\mu$ is 
centered between bands - thus the material is in insulator state. However, 
for $\bar{n}=\frac{1}{2}$, as an example, only half of the lower band is 
filled and this corresponds to the case of protonic conductor that occurs 
for example in superionic phases of superprotonic crystals.

We will discuss afterwards the consequences of the proton-phonon coupling 
effect focusing on the analysis of the strong coupling regime.
Since the low-order perturbation theory is insufficient in this case,
we apply a canonical Lang-Firsov transformation
\begin{eqnarray}
U={\rm e}^{iS}, \quad S=\sum\limits_{l\nu} n_{l\nu} v_{l\nu}, \quad
v_{la}=-v_{l-1,b}=i\sum_q \frac{\tau_l(q)}{\hbar\omega (q)}(b_q-b_{-q}^+).
\label{lang_firsov}
\end{eqnarray}
We present the transformed Hamiltonian
\begin{eqnarray}
\tilde{H}={\rm e}^{iS} H {\rm e}^{-iS}=\tilde{H}_0+\tilde{H}_t, 
\end{eqnarray}
where
\begin{eqnarray}
\tilde{H}_0&=&-\tilde{\mu}\sum\limits_{l\nu}n_{l\nu}+
E_0 \sum\limits_{l} n_{la}n_{l-1,b} + 
\sum\limits_q \hbar\omega(q) b_q^+ b_q, 
\\
\tilde{H}_t&=&\Omega_T \sum\limits_{l} (\Phi_T^l c_{la}^+ c_{lb}+
\Phi_T^{l*} c_{lb}^+ c_{la})+
\Omega_R \sum\limits_{l} (\Phi_R^l c_{la}^+ c_{l-1,b}+ 
\Phi_R^{l*} c_{l-1,b}^+ c_{la}). \nonumber
\end{eqnarray}
Here $\tilde{\mu}=\mu+E_0$ is the proton chemical potential which is 
renormalized due to the lattice polarization and the formation of the
protonic polaron as the result.
The notation $E_0=\sum\limits_q {|\tau_l(q)|^2}/{\hbar\omega(q)}=
(\hbar g)^2/2M(\hbar\omega_0)^2$ denotes the binding 
energy of the small polaron which forms in the strong proton-phonon
coupling regime and the phonon operators  
\begin{eqnarray}
\Phi_T^{l}={\rm e}^{-\sum\limits_q \frac{\tau_l(q)+\tau_{l+1}(q)}
{\hbar\omega (q)}(b_q-b_{-q}^+)}, \quad
\Phi_R^{l}={\rm e}^{-2\sum\limits_q \frac{\tau_l(q)}
{\hbar\omega (q)}(b_q-b_{-q}^+)},
\end{eqnarray}
give rise to the protonic polaron band narrowing factors with their thermal 
averages given by
\begin{eqnarray}
\langle \Phi_T^{l} \rangle={\rm e}^
{-\sum\limits_q {|\tau_l(q)|^2}{\rm cth} \frac{1}{2}\beta \hbar\omega (q)
/{(\hbar\omega(q))^2}}, \quad
\langle \Phi_R^{l} \rangle={\rm e}^
{-2\sum\limits_q {|\tau_l(q)|^2}{\rm cth} \frac{1}{2}\beta \hbar\omega (q)
/{(\hbar\omega(q))^2}}=\langle \Phi_T^{l} \rangle^2.
\end{eqnarray}

Furthermore we consider the case of $\bar{n}=1$ with the filled lower proton
band which corresponds to the insulator state of the system.

\section{CURRENT OPERATOR AND OPTICAL CONDUCTIVITY}

To analyze the protonic conductivity along the chain direction, we should 
evaluate the following expression in the framework of the linear response 
Kubo theory
\cite{kubo1}
\begin{eqnarray}
\sigma(\omega,T)=\frac{1}{Nd}\int_0^{\infty} dt 
\exp [i(\omega+i\varepsilon)t] \int_0^{\beta} d\lambda 
\langle J(t-i\hbar\lambda)J(0) \rangle, \label{kubo}
\end{eqnarray}
where the proton current operator
$J=\frac{e}{i\hbar}[H,x]=J_T+J_R$ ($x=\sum\limits_{l\nu} 
n_{l\nu}r_{l\nu}$ is the proton polarization operator and $r_{l\nu}$
is the proton coordinate in the $\nu$th position of the $l$th bond).
After the transformation (\ref{lang_firsov}) the parts $J_T$ and $J_R$
which represent the intrabond and interbond proton transfer respectively,
have the following form
\begin{eqnarray}
\tilde{J}_T&=&\frac{e \Omega_T}{i\hbar} R_{ab}\sum\limits_{l} (\Phi_T^l 
c_{la}^+ c_{lb}-\Phi_T^{l*} c_{lb}^+ c_{la}), \\
\tilde{J}_R&=&\frac{e \Omega_R}{i\hbar} R_{r} \sum\limits_{l} 
(\Phi_R^{l*} c_{l-1,b}^+ c_{la}-\Phi_R^l c_{la}^+ c_{l-1,b}), \nonumber
\end{eqnarray}
where $R_{ab}$ is the separation between the potential wells within 
the H-bond and $R_{r}$ is the distance between the nearest proton 
positions on the neighbouring bonds adjacent to the ionic group.
In this case the polaronic bandwidths 
$\Delta \varepsilon=\tilde{\Omega}_T+\tilde{\Omega}_R-
|\tilde{\Omega}_T-\tilde{\Omega}_R|$ 
($\tilde{\Omega}_T=\Omega_T \langle \Phi_T^{l} \rangle$,
$\tilde{\Omega}_R=\Omega_R \langle \Phi_R^{l} \rangle$) narrow exponentially
with temperature increase due to the proton-phonon scattering processes.
Because of this, the polaronic bands degenerate practically to localized 
states and the proton migration proceeds as the hopping of the protonic
polaron between localized positions in the lattice at high temperatures.

It is well known that the total polaronic conductivity can be represented as 
a sum of two parts\cite{polarons} : $\sigma=\sigma_{coh}+\sigma_{hop}$. The 
band conductivity $\sigma_{coh}$ dominates at low temperatures, whereas the 
hopping term $\sigma_{hop}$ makes the main contribution at high 
temperatures, when the bandwidths are small and the particles become trapped 
at individual sites due to phonon interactions.
We consider first the protonic low-temperature band conductivity. Since the 
scattering caused by phonon interactions can be neglected in this case, the 
proton transport is assumed to be the motion in the thermally averaged 
phonon field. Then evaluation of the correlation functions in (\ref{kubo})
using the Wick's theorem yields the following expression for the 
real part of $\sigma_{coh}$
\begin{eqnarray}
\sigma_{coh}=\sigma_{coh}^0+\sigma_{coh}^{ab},
\end{eqnarray}
where the Drude-type term can be written as
\begin{eqnarray}
\sigma_{coh}^0=-\pi\frac{e^2\beta}{4\hbar^2 Nd} \sum_{k, \nu} 
\frac{(\tilde{\Omega}_k \tilde{t}_k^* -\tilde{\Omega}_k^* 
\tilde{t}_k)^2}{|\tilde{t}_k|^2}
\langle \tilde{n}_{k\nu} \rangle (1-\langle \tilde{n}_{k\nu} \rangle )
\delta (\omega)
\end{eqnarray}
with 
$\tilde{\Omega}_k=\tilde{\Omega}_T R_{ab}-\tilde{\Omega}_R R_r 
{\rm e}^{-ikd}$, $\tilde{t}_k=\tilde{\Omega}_T+\tilde{\Omega}_R {\rm 
e}^{-ikd}$, and $\sigma_{coh}^{ab}$ is given by
\begin{eqnarray}
\sigma_{coh}^{ab}=\frac{e^2}{\hbar d}
\frac{ {\rm sh} \beta \hbar\omega/2}{(\hbar\omega)^2}
\frac{1}{ {\rm ch} \beta \tilde{\mu} +
{\rm ch} \beta\hbar\omega/2}
\frac{[(\tilde{\Omega}_T^2-\tilde{\Omega}_R^2)(R_{ab}+R_r) +
(\hbar\omega/2)^2 (R_{ab}-R_r)]^2}
{\sqrt{(2\tilde{\Omega}_T \tilde{\Omega}_R)^2-
[(\hbar\omega/2)^2-(\tilde{\Omega}_T + \tilde{\Omega}_R)^2)]^2}}
\label{rsigma_coh_ab}
\end{eqnarray}
Here the notations $\langle \tilde{n}_{k\nu} \rangle=
[1+\exp(\beta(\pm |\tilde{t}_k|-\tilde{\mu}))]^{-1}$ 
denote thermally averaged Fermi distributed polaronic band occupancies. 
We see that the first term $\sigma_{coh}^0$ corresponds to the
intraband transfer processes giving rise to the 
absorption peak at $\omega=0$, whereas 
$\sigma_{coh}^{ab}$ is associated with the interband transfer 
processes and leads to appearance of the van-Hove-type singularities in the 
absorption at $\hbar\omega=\pm (\tilde{\Omega}_T + \tilde{\Omega}_R)$ and
$\hbar\omega=\pm |\tilde{\Omega}_T - \tilde{\Omega}_R|$.

\section{THE PHONON-ACTIVATED PROTON TRANSPORT}

Furthermore, we will focus on the analysis of the polaronic transport
which corresponds to phonon-activated hopping, because of the dominant 
contribution of this part to the total protonic conductivity at high 
temperatures. 

We consider the following correlator in the conductivity relation 
(\ref{kubo}) which characterizes the hopping transport
\begin{eqnarray}
K_{hop}=\langle \tilde{J}(z) 
\tilde{J}(0)\rangle_{hop}=K_{hop}^T+K_{hop}^R+K_{hop}^{TR}
\end{eqnarray}
where $z=t-i\hbar\lambda$ and
\begin{eqnarray}
K_{hop}^T=&&\frac{e^2 \Omega_T^2}{\hbar^2} R_{ab}^2 \sum_{ll'}
\left( \langle\tilde{\Phi}_T^l(z) c_{la}^+(z)c_{lb}(z) 
\tilde{\Phi}_T^{l'*} c_{l'b}^+ c_{l'a} \rangle +
\langle\tilde{\Phi}_T^{l*}(z) c_{lb}^+(z)c_{la}(z) 
\tilde{\Phi}_T^{l'} c_{l'a}^+ c_{l'b} \rangle \right), \nonumber\\
K_{hop}^R=&&\frac{e^2 \Omega_R^2}{\hbar^2} R_{r}^2 \sum_{ll'}
\left( \langle\tilde{\Phi}_R^l(z) c_{l-1,b}^+(z)c_{la}(z) 
\tilde{\Phi}_R^{l'*} c_{l'a}^+ c_{l'-1,b} \rangle + \right.\nonumber\\
&&\left. \langle\tilde{\Phi}_R^{l*}(z) c_{la}^+(z)c_{l-1,b}(z) 
\tilde{\Phi}_R^{l'} c_{l'-1,b}^+ c_{l'a} \rangle \right), \label{K_hop}\\
K_{hop}^{TR}=&&-\frac{e^2 \Omega_0\Omega_R}{\hbar^2}R_{ab} R_{r} \sum_{ll'}
\left( \langle\tilde{\Phi}_T^l(z) c_{la}^+(z)c_{lb}(z) 
\tilde{\Phi}_R^{l'*} c_{l'-1,b}^+ c_{l'a} \rangle + \right.\nonumber\\
&& \langle\tilde{\Phi}_T^{l*}(z) c_{lb}^+(z)c_{la}(z) 
\tilde{\Phi}_R^{l'} c_{l'a}^+ c_{l'-1,b} \rangle +
\langle\tilde{\Phi}_R^{l*}(z) c_{l-1,b}^+(z)c_{la}(z) 
\tilde{\Phi}_T^{l'} c_{l'a}^+ c_{l'b} \rangle + \nonumber\\
&& \left. \langle\tilde{\Phi}_R^{l}(z) c_{la}^+(z)c_{l-1,b}(z) 
\tilde{\Phi}_T^{l'*} c_{l'b}^+ c_{l'a} \rangle \right) \nonumber
\end{eqnarray}
Here the operators $\tilde{\Phi}_{\alpha}^l=\Phi_{\alpha}^l-\langle 
\Phi_{\alpha}^l \rangle$ ($\alpha=T,R$) correspond to the phonon scattering 
processes which accompany the proton transfer. We perform evaluation of 
(\ref{K_hop}) using the Hamiltonian with averaged values of the transfer 
integrals $\tilde{\Omega}_{\alpha}$. This allows us to decouple the 
proton-phonon correlation functions 
\begin{eqnarray}
\langle \tilde{\Phi}_{\alpha}^m(z) c_{l\nu}^+(z)c_{l'\nu'}(z) 
\tilde{\Phi}_{\alpha_1}^{m_1} c_{l_1,\nu_1}^+ c_{l_1'\nu_1'} \rangle 
\rightarrow
\langle \tilde{\Phi}_{\alpha}^m(z) \tilde{\Phi}_{\alpha_1}^{m_1}\rangle 
\langle c_{l\nu}^+(z)c_{l'\nu'}(z) c_{l_1,\nu_1}^+ c_{l_1'\nu_1'} \rangle 
\end{eqnarray}
into proton and phonon parts which can be evaluated separately.
The processes that we are interested in are going on at considerably
high temperatures and in the strong proton-phonon coupling regime
(we will see below that the strong polaron effect actually occurs in these 
systems). This allows us to assume
\begin{eqnarray}
\sum\limits_q \frac{|\tau_l(q)|^2}
{(\hbar\omega(q))^2 {\rm sh} \beta \hbar\omega(q)/2} >>1.
\end{eqnarray}
We take a dispersionless approximation for the phonon 
frequency: $\omega(q)=\omega_0$.
In this case $E_0=(\hbar g)^2/2M(\hbar\omega_0)^2$ and  
evaluation of the phonon correlation functions yields
\begin{eqnarray}
\psi_{TT}=&&\langle \tilde{\Phi}_T^l(z) \tilde{\Phi}_T^{l'*}\rangle= 
\langle \tilde{\Phi}_T^{l*}(z) \tilde{\Phi}_T^{l'}\rangle=
{\rm e}^{-\beta E_0/2} {\rm e}^{-(z+i\hbar\beta/2)^2/4\tilde{\tau}_T^2}
\delta_{ll'}+ \nonumber\\
&&{\rm e}^{-\beta E_0/4} {\rm e}^{-\frac{E_0}{\hbar \omega_0}
{\rm cth} \beta \hbar \omega_0/2}
{\rm e}^{-(z+i\hbar\beta/2)^2/4\tilde{\tau}_T'^2} \delta_{l,l'\pm 1},
\label{psi} \\
\psi_{RR}=&&\langle \tilde{\Phi}_R^l(z) \tilde{\Phi}_R^{l'*}\rangle= 
\langle \tilde{\Phi}_R^{l*}(z) \tilde{\Phi}_R^{l'}\rangle=
{\rm e}^{-\beta E_0} {\rm e}^{-(z+i\hbar\beta/2)^2/4\tilde{\tau}_R^2}
\delta_{ll'}, \nonumber\\
\psi_{TR}=&&\langle \tilde{\Phi}_T^l(z) \tilde{\Phi}_R^{l'*}\rangle= 
\langle \tilde{\Phi}_T^{l*}(z) \tilde{\Phi}_R^{l'}\rangle=
\langle \tilde{\Phi}_R^l(z) \tilde{\Phi}_T^{l'*}\rangle= 
\langle \tilde{\Phi}_R^{l*}(z) \tilde{\Phi}_T^{l'}\rangle=
{\rm e}^{-\beta E_0/2} \times \nonumber\\
&& {\rm e}^{-\frac{E_0}{\hbar \omega_0}{\rm cth} \beta \hbar \omega_0/2} 
{\rm e}^{-(z+i\hbar\beta/2)^2/4\tilde{\tau}_T^2}
(\delta_{ll'}+\delta_{l,l'\pm 1}) \nonumber
\end{eqnarray}
The parameters introduced in (\ref{psi}) 
\begin{eqnarray}
\tilde{\tau}_T^2=\frac{\beta \hbar^2}{8E_0}, & \quad
\displaystyle{\tilde{\tau}_R^2=\frac{\beta \hbar^2}{16E_0},} & \quad
\tilde{\tau}_T'^2=\frac{\beta \hbar^2}{4E_0}
\end{eqnarray}
characterize the average hopping time lengths between two localized 
positions in the chain (the hopping time within H-bond, the interbond 
reorientational hopping time and the time for pair
polaron transfer within the neighbouring bonds respectively).

Next we consider the proton part of the correlation functions. We introduce 
the $k$-representation for the Fermi operators
\begin{eqnarray*}
c_{l\nu}=\frac{1}{\sqrt{N}} \sum\limits_k c_{k\nu} {\rm e}^{ikld}
\end{eqnarray*}
and further perform the following unitary transformation
\begin{eqnarray*}
c_{ka}=\frac{1}{d_k} \left( [\tilde{t}_k-i|\tilde{t}_k|]\tilde{c}_{ka}-
[|\tilde{t}_k|+i\tilde{t}_k]\tilde{c}_{kb} \right),\\
c_{kb}=\frac{1}{d_k} \left( [|\tilde{t}_k|-i\tilde{t}_k^*]\tilde{c}_{ka}+
[\tilde{t}_k^*+i|\tilde{t}_k|]\tilde{c}_{kb} \right)
\end{eqnarray*}
where $d_k=2\sqrt{|\tilde{t}_k|} \sqrt{|\tilde{t}_k|-\Im \tilde{t}_k}$.
Then the polaron Hamiltonian can be reduced to the diagonal form
$\sum_{k\nu} \varepsilon_{k\nu} \tilde{n}_{k\nu}$ 
($\varepsilon_{k\nu}=-\tilde{\mu} \pm |\tilde{t}_k|$)
and the correlation functions 
$\langle \tilde{c}_{k\nu}^+(z) \tilde{c}_{k'\nu'}(z)
\tilde{c}_{k_1\nu_1}^+ \tilde{c}_{k_1'\nu_1'}\rangle$ can be evaluated using 
the Wick's theorem (in this case we use the Schr\"{o}dinger representation 
for the polaron operators $\tilde{c}_{k\nu}(z)={\rm e}^{-i 
\varepsilon_{k\nu}z/\hbar} \tilde{c}_{k\nu}$).

We use procedure proposed in \cite{reik} with the deformation of the 
integration contour in the complex plane for the integration of expressions 
(\ref{K_hop}) over $t$ and $\lambda$. Let us present the resulting 
expression for the real part of the conductivity that will be analyzed 
afterwards
\begin{eqnarray}
&&\sigma_{hop}(\omega,T)=\sigma_{hop}^0(\omega,T)+
\sum\limits_{\nu \nu'}\sigma_{hop}^{\nu,\nu'}(\omega,T), \label{sigma_hop}\\
&&\sigma_{hop}^0(\omega,T)=\frac{2e^2\sqrt{\pi}}{dN^2\hbar^2} 
\frac{{\rm sh}\beta\hbar\omega/2}{\hbar\omega/2} \sum_{kk'}
\left\{ (C_T {\rm e}^{-\tilde{\tau}_T^2\omega^2}+  
C_T' {\rm e}^{-\tilde{\tau}_T'^2\omega^2}) \Re (\tilde{t}_k^* 
\tilde{t}_{k'})+ \right.\nonumber\\
&& \left. C_R {\rm e}^{-\tilde{\tau}_R^2\omega^2}
\Re (\tilde{t}_k^* \tilde{t}_{k'}{\rm e}^{-i(k-k')d})+
C_{TR} {\rm e}^{-\tilde{\tau}_T^2\omega^2}
\Re (\tilde{t}_k^* \tilde{t}_{k'}[{\rm e}^{ik'd}+{\rm e}^{-ikd}])
\right\} \frac{(\langle \tilde{n}_{ka}\rangle- \langle\tilde{n}_{kb} 
\rangle) (\langle \tilde{n}_{k'a}\rangle- \langle\tilde{n}_{k'b} \rangle)}
{|\tilde{t}_k| |\tilde{t}_{k'}|} \nonumber \\
&&\sigma_{hop}^{\nu,\nu'}(\omega,T)=\frac{2e^2\sqrt{\pi}}{dN^2\hbar^2} 
\frac{{\rm sh}\beta\hbar\omega/2}{\hbar\omega/2} \sum_{kk'}
\left\{ C_T {\rm e}^{-\tilde{\tau}_T^2\tilde{\omega}^2}+  
C_T' {\rm e}^{-\tilde{\tau}_T'^2\tilde{\omega}^2}\cos (k-k')d + 
\right.\nonumber\\
&& \left. C_R {\rm e}^{-\tilde{\tau}_R^2\tilde{\omega}^2}
+C_{TR} {\rm e}^{-\tilde{\tau}_T^2\tilde{\omega}^2}
(\cos kd +\cos k'd) \right\} 
{\rm e}^{\beta (\varepsilon_{k\nu}-\varepsilon_{k'\nu'})/2}
\langle \tilde{n}_{k\nu}\rangle 
(1-\langle\tilde{n}_{k'\nu'} \rangle). \nonumber 
\end{eqnarray}
Here 
$\tilde{\omega}=\omega+\frac{\varepsilon_{k\nu}-\varepsilon_{k'\nu'}}{\hbar}$ 
and
\begin{eqnarray*}
C_T=\Omega_T^2 R_{ab}^2 \tilde{\tau}_T {\rm e}^{-\beta E_0/2}, \quad
C_T'=2\Omega_T^2 R_{ab}^2 \tilde{\tau}_T' {\rm e}^{-\beta E_0/4}
{\rm e}^{-\frac{E_0}{\hbar\omega_0} {\rm cth} \beta\hbar\omega_0/2}, \\ 
C_R=\Omega_R^2 R_{r}^2 \tilde{\tau}_R {\rm e}^{-\beta E_0}, \quad
C_{TR}=-2\Omega_T \Omega_R R_{ab} R_{r} \tilde{\tau}_T 
{\rm e}^{-\beta E_0/2}
{\rm e}^{-\frac{E_0}{\hbar\omega_0} {\rm cth} \beta\hbar\omega_0/2}. 
\end{eqnarray*}
The first term in (\ref{sigma_hop}) appears in the result of pairing
of the operators $\tilde{c}_{k\nu}^+$ with $\tilde{c}_{k'\nu'}$
in the proton correlation functions with equal values of time argument, 
whereas another terms $\sigma_{hop}^{\nu,\nu'}$ arise due to the pairing
of such operators with different time arguments. 

\begin{figure}[htbp]
\epsfxsize=7.cm
\epsfysize=7.cm
\centerline{\epsffile{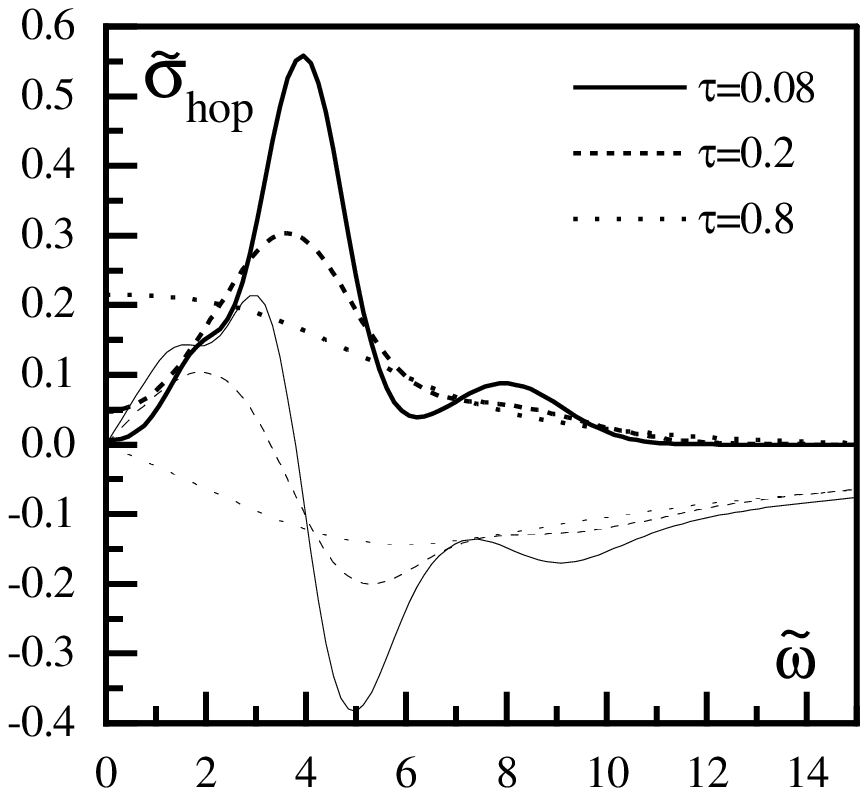}}
\caption{}
\label{fig2}
\end{figure}

Let us discuss first the 
frequency dependence of the hopping conductivity. From expression 
(\ref{sigma_hop}) it follows that $\sigma_{hop}$ has several peaks which 
point to the important contribution of various elementary dynamical 
processes existing in the system. It is possible to estimate analytically 
in the limit $\beta\hbar\omega>>1$ (in the high frequency range or at lower 
temperatures) the characteristic frequency values at which the conductivity 
maxima exist.
The first maximum at $\omega_1 \sim \frac{\beta\hbar}{4\tilde{\tau}_T'^2}$ 
corresponds to the pair phonon-activated hopping within the neighbouring 
H-bonds. Another maximum at $\omega_2 
\sim \frac{\beta\hbar}{4\tilde{\tau}_T^2}$ arises due to the intrabond 
polaron dynamics, whereas the peak at 
$\omega_3 \sim \frac{\beta\hbar}{4\tilde{\tau}_R^2}$ is associated with the 
reorientational hopping of the protonic polaron between bonds. The frequency 
dependencies of the real and imagine parts of the hopping conductivity are 
shown in Figs.~\ref{fig2}, 3 for different temperatures and values of 
polaron binding energy $E_0$. We observe three distinct peaks at 
sufficiently low temperatures, which broaden significantly when 
temperature increase. Finally, at a considerably high temperatures the 
behavior $\sigma_{hop}(\omega)$ has the Drude-type character with maximum 
at $\omega=0$ and smoothing decrease with increasing frequency.

The two-stage transport mechanism manifests itself in the two peaks at
$\omega=\omega_2$ and $\omega=\omega_3$ observed in the frequency
dependence of the optical conductivity which shift to the high-frequency 
region when the polaron binding energy $E_0$ increases (see Fig.~3). With 
the increase of $E_0$ the conductivity lowers and the maxima broaden due to 
the stronger localization of protonic polaron in one of the potential well 
of the hydrogen bond. We note that the values of the hopping integrals 
$\tilde{\Omega}_T$ and $\tilde{\Omega}_R$ practically do not influence on 
the positioning of these maxima but affect strongly the redistribution of 
the maxima heights.

Figure~4 shows the comparison of theoretical conductivity obtained for 
lithium hydrazinium sulphate Li(N$_2$H$_5$)SO$_4$ (LHS) using 
(\ref{sigma_hop}) with experimentally measured values \cite{salman}
at the frequency 500~KHz. This compound consists of the chains of hydrogen 
bonded hydrazinium ions running along the [001] direction and can be 
considered as the example of quasi-one-dimensional protonic conductor
\cite{padmanabhan}. In this case we obtain $\hbar \omega_0=800$~cm$^{-1}$ 
and $E_0=0.98$~eV. According to our results, the dominant contribution to 
the conductivity in this frequency range arises from the intrabond hopping 
with $\Omega_T=0.04$~eV. The relation $E_0>2\Omega_T$ follows immediately 
from the values of our parameters and is valid for the small-polaron regime
\cite{polarons}.

Let us analyze the theoretical hopping conductivity in 
a wide temperature interval for the various system parameters. 
Since the proton is localized in the chain due to the strong polaronic 
effect, we expect intuitively that change in an external control parameters 
may have a determining influence on the proton state and lead to its 
delocalization. In particular, it is of prime importance to analyze in more 
details the influence of temperature. On the one hand, it is well known that 
the effect of temperature is dramatic in the systems which undergo 
the transitions to superprotonic phases, which accompanied by significantly 
increase of the protonic conductivity. In this case the proton state changes 
especially drastically: in the low-temperature ordered phases the protons 
are localized in well-defined positions of the lattice and can migrate with 
sufficiently high activation energy ($\sim 1$~eV) due to the strong coupling 
with displacements of ionic groups participating in hydrogen bond formation, 
whereas in the superionic phases the transport phenomenon 
is connected with dynamically disorder of the hydrogen bond network, 
resulting in an increase of the number of possible positions for protons 
which can hop between them with much lower activation energy ($\sim 0.1$~eV).
On the other hand, the influence of temperature on the localized state of 
protonic polaron can be especially important in a variety of protonic 
conductors, both in inorganic systems and biological membranes.
Because of this, we consider the temperature dependence of the hopping 
conductivity at the given frequency of external field.

The real part of the $\sigma_{hop}$ as a function of temperature has 
peculiarities which arise due to delocalization effects. We consider several
characteristic frequency regions: $\omega \sim \omega_2$ (here the intrabond 
hopping contributes mainly to the total conductivity), $\omega \sim 
\omega_3$ (when the reorientational interbond hopping dominates) and 
$\omega=0$ (static dc conductivity). The effect of two-stage transport
process shows itself most clearly in the high-frequency region.
The first case is shown in Fig.~5. The two maxima are observed which shift 
slightly towards the higher temperatures when $E_0$ increases. It is easy to 
understand the origin of this nonmonotonic behavior of $\sigma_{hop}$ from 
Fig.~6, where the temperature dependence of $\sigma_{hop}$ is compared with 
two extreme cases that one of the transport component (interbond or 
intrabond) is eliminated. Now we can conclude immediately that the first 
sharp increase of $\sigma_{hop}$ with further rapid lowering originates 
from the intrabond hopping, whereas the second smoother maximum on higher 
temperature interval corresponds to the interbond contribution. The raising 
of temperature suppresses the polaron localization effect: the thermal 
proton dynamics leads to the detrapping of protonic polaron which 
accompanied by drastically increase of $\sigma_{hop}$. However, with the 
further temperature increasing the conductivity lowers again 
due to the multiphonon scattering processes suggesting
that the system undergoes a transition to the metallic-type state.
Therefore, we can conclude that in our 
system the possible transition from the insulator state has a two-step
character: first the protonic polaron is delocalized within the H-bond, and 
only with further temperature increase it becomes delocalized between 
bonds. The situation changes alternatingly in the vicinity of $\omega \sim 
\omega_3$
(see Figs.~7 and 8). We see that in this case first the tendency for 
interbond delocalization occurs, whereas the intrabond contribution 
dominates at higher temperatures. 
The last case $\sigma_{hop}(\omega=0)$ is shown in Fig.~9. for different
relations between $\Omega_T$ and $\Omega_R$. We see the single maximum 
which corresponded to the transition to metallic-type state appeared due to 
the dominant contribution of reorientations (when $\Omega_R> \Omega_T$, 
Fig.~9(a)) or intrabond hopping (when $\Omega_T> \Omega_R$, 
Fig.~9(b)). In the case $\Omega_T \sim \Omega_R$ (Fig.~9(c)) the both parts 
contribute practically equally to the transition occurrence.
It is worth noting that despite of the single smooth peak, we can also
distinguish two steps in the dc conductivity behavior which becomes clear 
after comparison of our curves with the two extreme cases ($\Omega_R=0$ and 
$\Omega_T=0$). We can conclude that the protonic polarons become 
delocalized between bonds at lower temperatures, and the intrabond 
delocalization proceeds on further heating. This effect can be explained by 
the fact that the reorientational hopping has higher activation energy than 
the interbond motion (see expressions (\ref{sigma_hop}) for the 
conductivity) and thus the first type of hopping has the tendency to damp 
at lower temperatures due to the multiphonon scattering than the second 
type of hopping. Thus our model allows us a possibility to 
analyze thoroughly the effect of either of two transport stages on the total 
conductivity picture that can be verified by experimental measurements.

\section{CONCLUSIONS}

In the present work the protonic conductivity 
of the quantum quasi-one-dimensional hydrogen-bonded chain
has been studied in the framework of the two-stage orientational-tunneling 
model. The interaction of protons with the optical stretching mode is 
considered. The expression for the conductivity coefficient as a function
of temperature and frequency is obtained using the Kubo formula for 
various values of hopping parameters and polaron binding energy.
The evaluated hopping conductivity is in good agreement with the 
experimentally measured values for LHS system. This allows us to conclude 
that the strong proton-phonon coupling regime exists in this type of 
systems. The analysis of theoretically obtained temperature dependencies of 
$\sigma_{hop}$ in wide temperature interval revealed that the transition 
from the insulator state (increase of $\sigma$ with increase of T) to the 
metal-type state (decrease of $\sigma$ with increase of T) occurs in system 
on heating. The specific two-stage Grotthuss transport mechanism manifests 
itself in the two-step character of the transition as well as in the two 
distinct peaks of dynamical conductivity. The obtained theoretical results 
allow us to believe that the proposed theory should also apply to other 
hydrogen bonded systems and to investigate the consequences of assumptions 
of the Grotthuss mechanism for proton migration in hydrogen bonded 
compounds.

\section*{Acknowledgements}

This work is partially supported by INTAS Grant No.~95-0133.

\newpage
\section*{Figure captions}

\noindent
{\bf Figure 1.} (a) Zig-zag hydrogen-bonded chain, 
arrows indicate the possible path of proton migration along the chain.
(b) Simplified model chains, the in-phase displacements
of ionic groups identified by the dashed arrows.\\
{\bf Figure 2.} Frequency dependencies of the hopping conductivity
$\tilde{\sigma}_{hop}(\tilde{\omega})=\sigma_{hop}/c_0$ 
($c_0=2\frac{e^2}{d\hbar} r_0^2$, $\tilde{\omega}=\omega/\omega_0$) for 
$\tilde{E}_0=E_0/\hbar\omega_0=2$ and different
temperatures $\tau=kT/\hbar\omega_0$; $R_{ab}/r_0=1.3$, $R_{r}/r_0=1.6$.
$\omega_T=\tilde{\Omega}_T/\hbar\omega_0=0.8$ and 
$\omega_R=\tilde{\Omega}_R/\hbar\omega_0=0.5$ are reduced hopping amplitudes.
The real and imagine parts are indicated by bold and thin curves 
respectively.\\
{\bf Figure 3.} Frequency dependencies of $\Re\tilde{\sigma}_{hop}$ 
at $\tau=0.25$ for different values of $\tilde{E}_0$ and hopping 
amplitudes.\\
{\bf Figure 4.} Comparison of the temperature dependencies of the protonic 
conductivity, measured for the crystal LHS and evaluated using 
(\ref{sigma_hop}).\\
{\bf Figure 5.} Temperature dependencies of $\Re\tilde{\sigma}_{hop}$ 
for different values of $E_0$
at $\tilde{\omega}=5.5$, $\omega_T=0.2$ and $\omega_R=0.6$.\\
{\bf Figure 6.} Comparison of $\Re\tilde{\sigma}_{hop}(T)$ for 
$\tilde{E}_0=2.5$, $\tilde{\omega}=5.5$, $\omega_T=0.2$ and 
$\omega_R=0.6$ with two extreme cases when the intrabond or interbond 
contribution is eliminated.\\
{\bf Figure 7.} Temperature dependencies of $\Re\tilde{\sigma}_{hop}$ 
for different values of $E_0$
at $\tilde{\omega}=8.7$, $\omega_T=0.6$ and $\omega_R=0.3$.\\
{\bf Figure 8.} Comparison of $\Re\tilde{\sigma}_{hop}(T)$ for 
$\tilde{E}_0=2.3$, $\tilde{\omega}=8.7$, $\omega_T=0.6$ and 
$\omega_R=0.3$ with two extreme cases when the intrabond or interbond 
contribution is eliminated.\\
{\bf Figure 9.} Temperature dependencies of dc 
hopping conductivity for different values of 
$\omega_T$ and $\omega_R$ at $\tilde{E}_0=2.5$.

\newpage

\null\vspace{0.2in}

\begin{figure}[htbp]
\epsfxsize=7.cm
\epsfysize=7.cm
\centerline{\epsffile{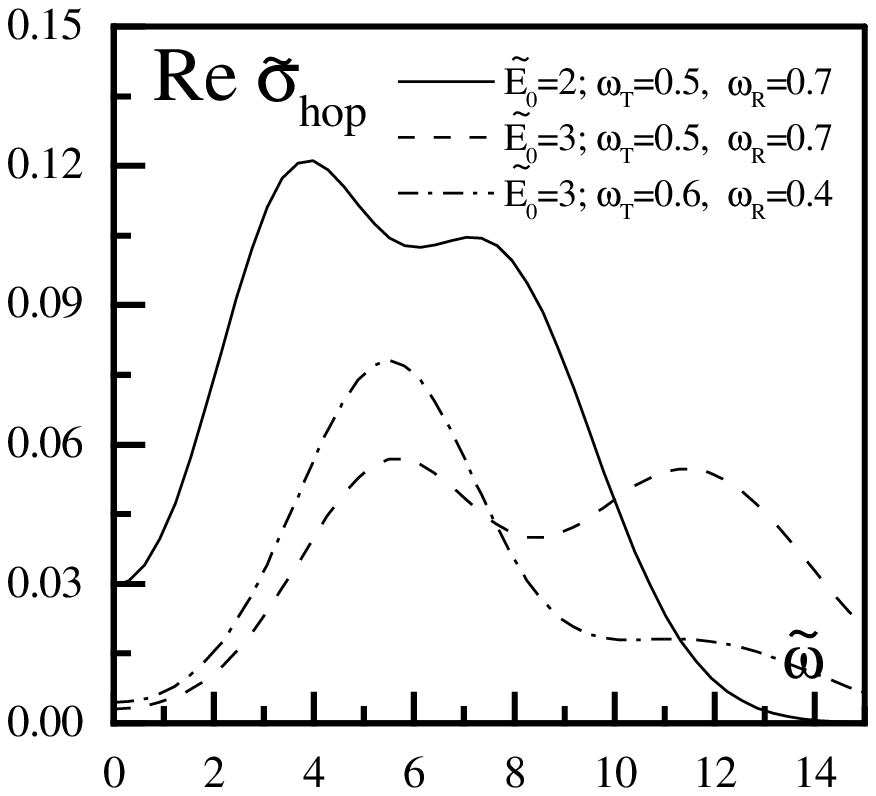}}
\caption{}
\label{fig3}
\end{figure}
\null\vspace{0.2in}

\begin{figure}[htbp]
\epsfxsize=7.cm
\epsfysize=7.cm
\centerline{\epsffile{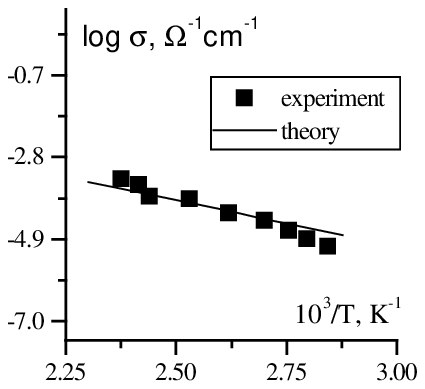}}
\caption{}
\label{fig4}
\end{figure}
\null\vspace{0.2in}

\begin{figure}[htbp]
\epsfxsize=7.cm
\epsfysize=7.cm
\centerline{\epsffile{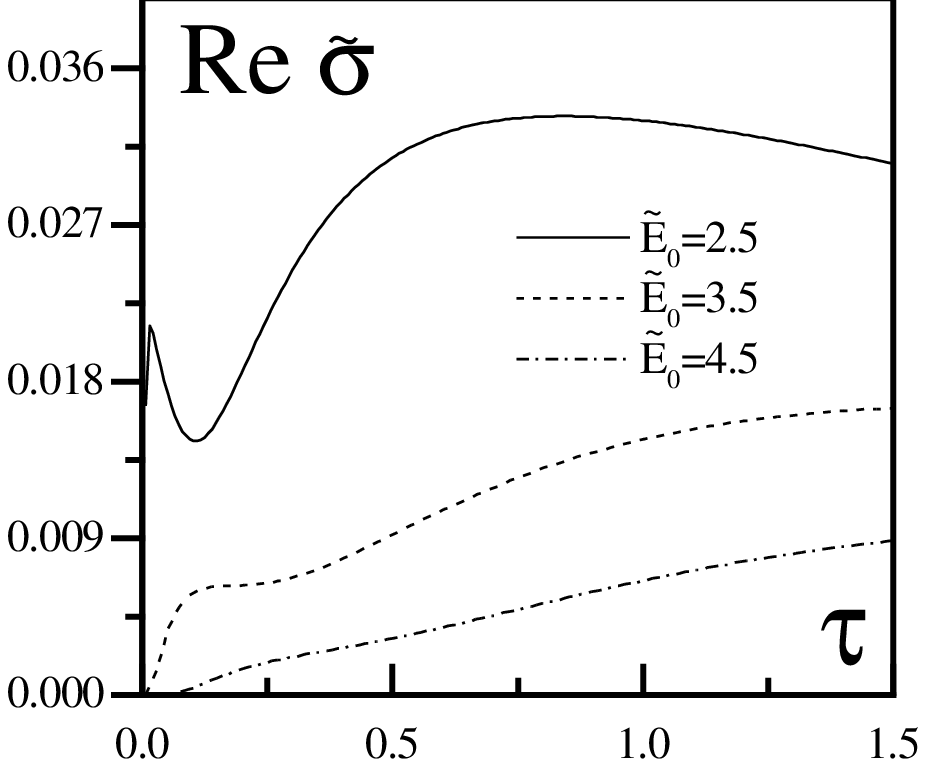}}
\caption{}
\label{fig5}
\end{figure}
\null\vspace{0.2in}

\begin{figure}[htbp]
\epsfxsize=7.cm
\epsfysize=7.cm
\centerline{\epsffile{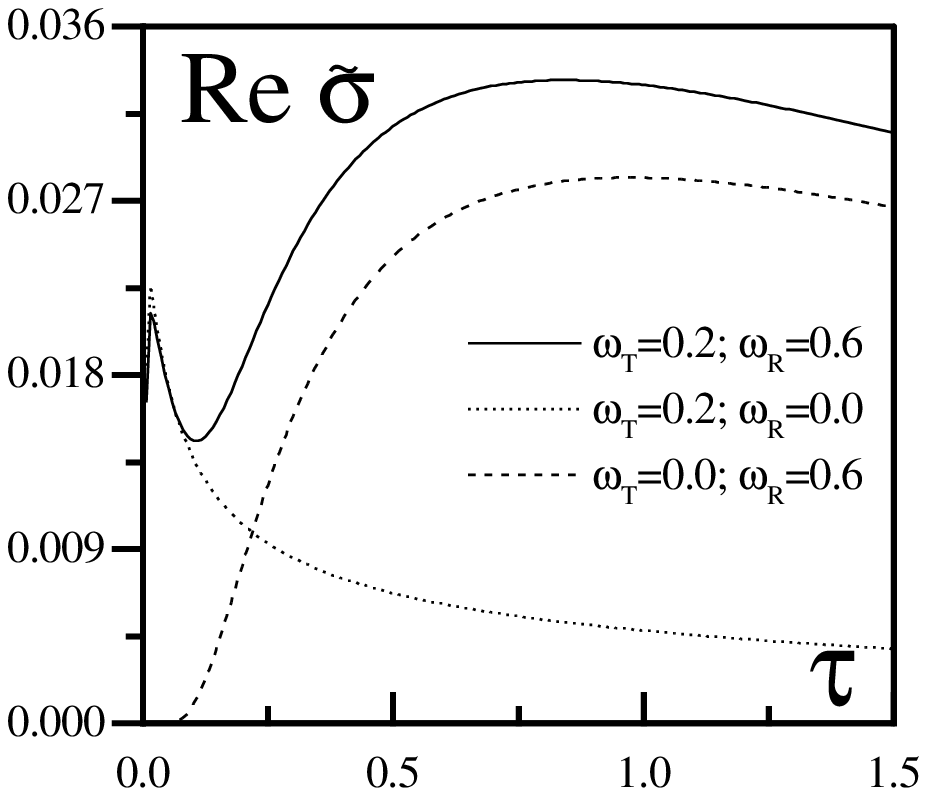}}
\caption{}
\label{fig6}
\end{figure}
\null\vspace{0.2in}

\begin{figure}[htbp]
\epsfxsize=7.cm
\epsfysize=7.cm
\centerline{\epsffile{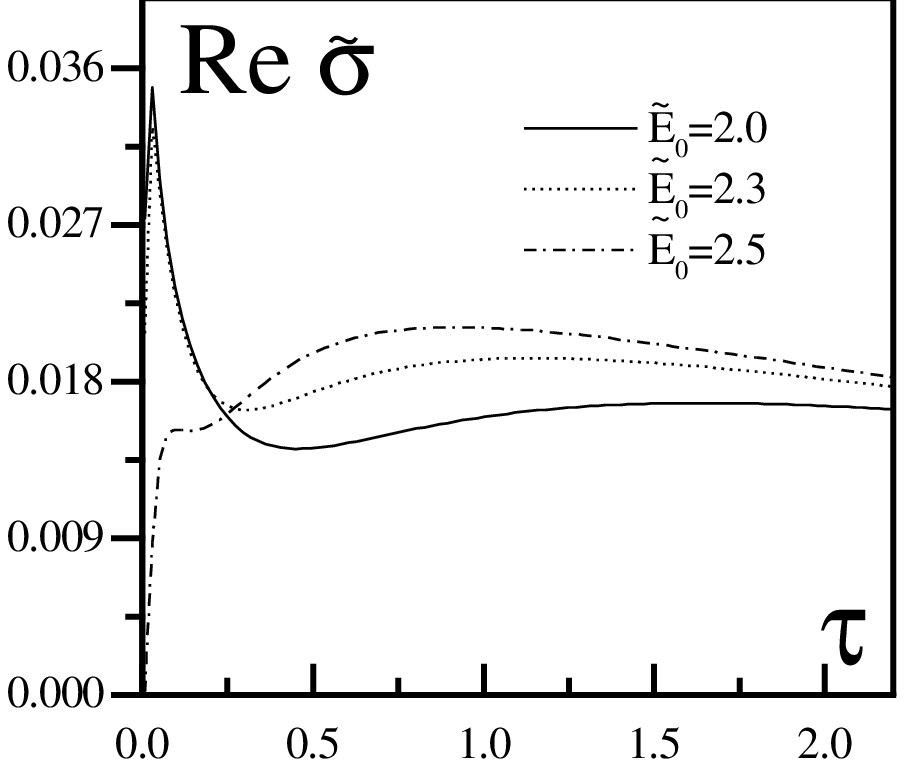}}
\caption{}
\label{fig7}
\end{figure}
\null\vspace{0.2in}

\begin{figure}[htbp]
\epsfxsize=7.cm
\epsfysize=7.cm
\centerline{\epsffile{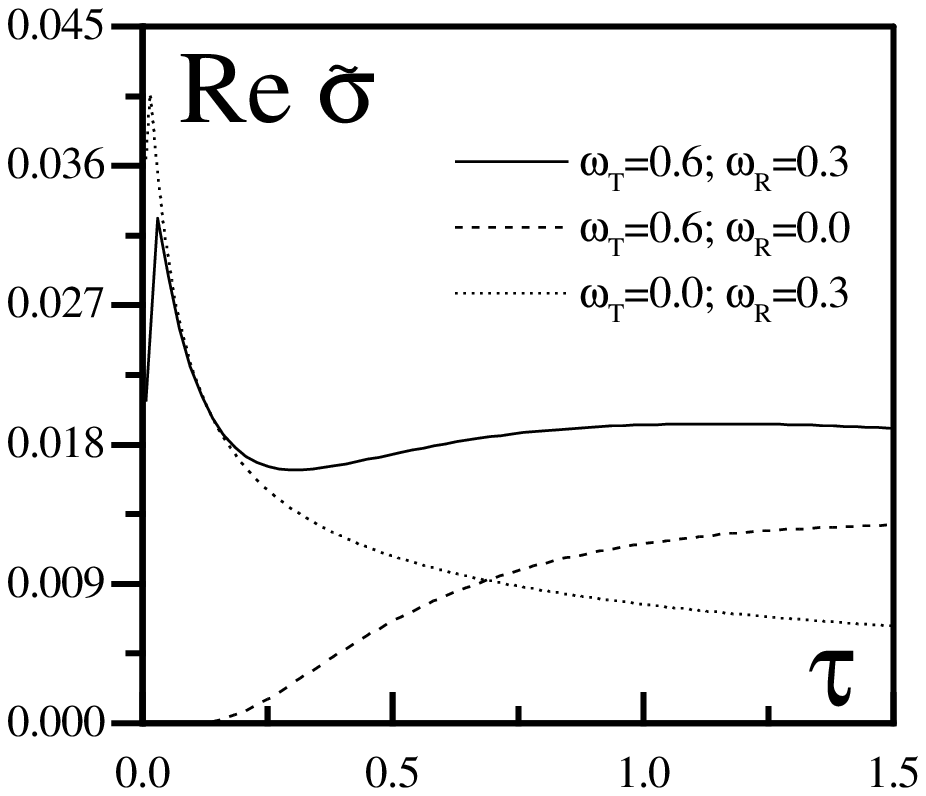}}
\caption{}
\label{fig8}
\end{figure}
\null\vspace{0.2in}

\begin{figure}[htbp]
\epsfxsize=7.cm
\epsfysize=7.cm
\centerline{\epsffile{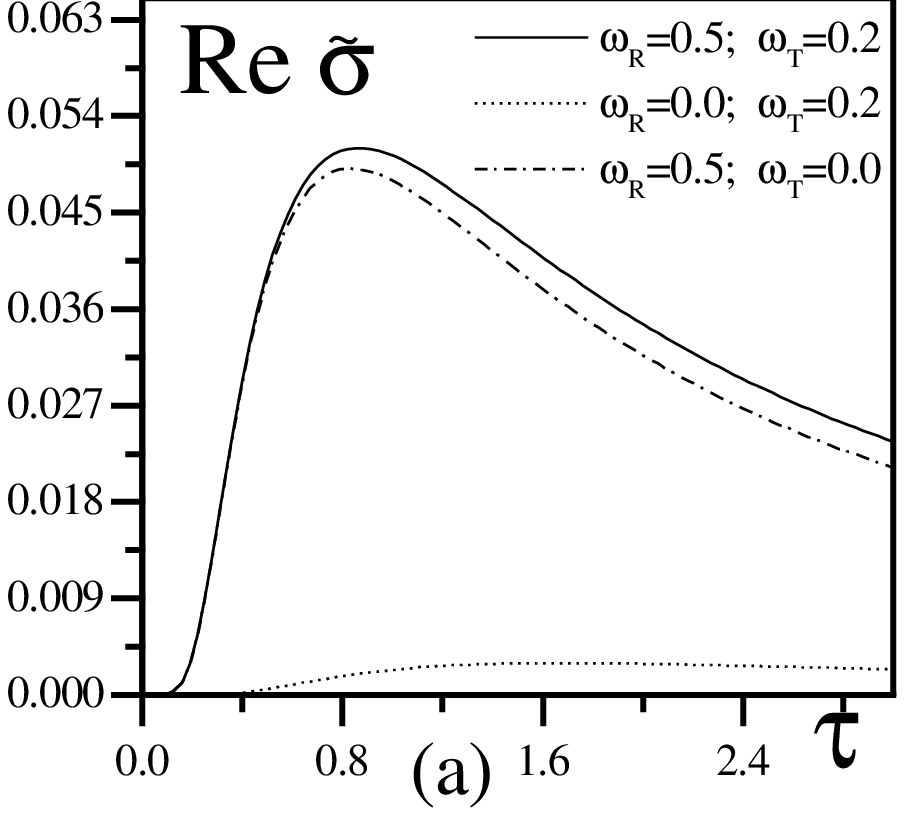}}
\epsfxsize=7.cm
\epsfysize=7.cm
\centerline{\epsffile{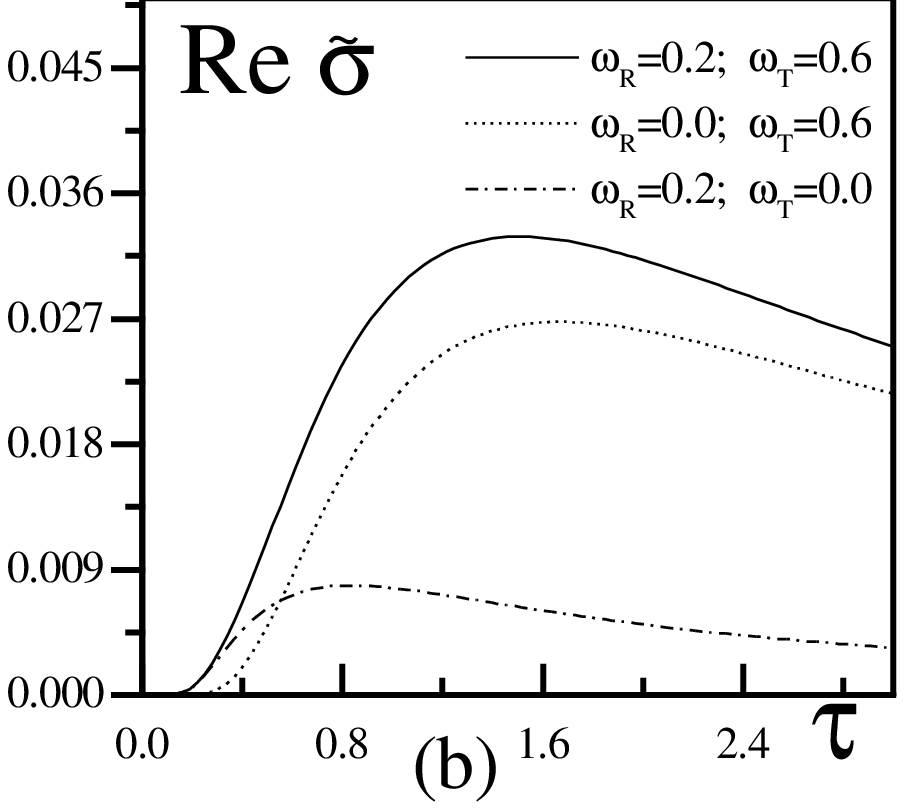}}
\epsfxsize=7.cm
\epsfysize=7.cm
\centerline{\epsffile{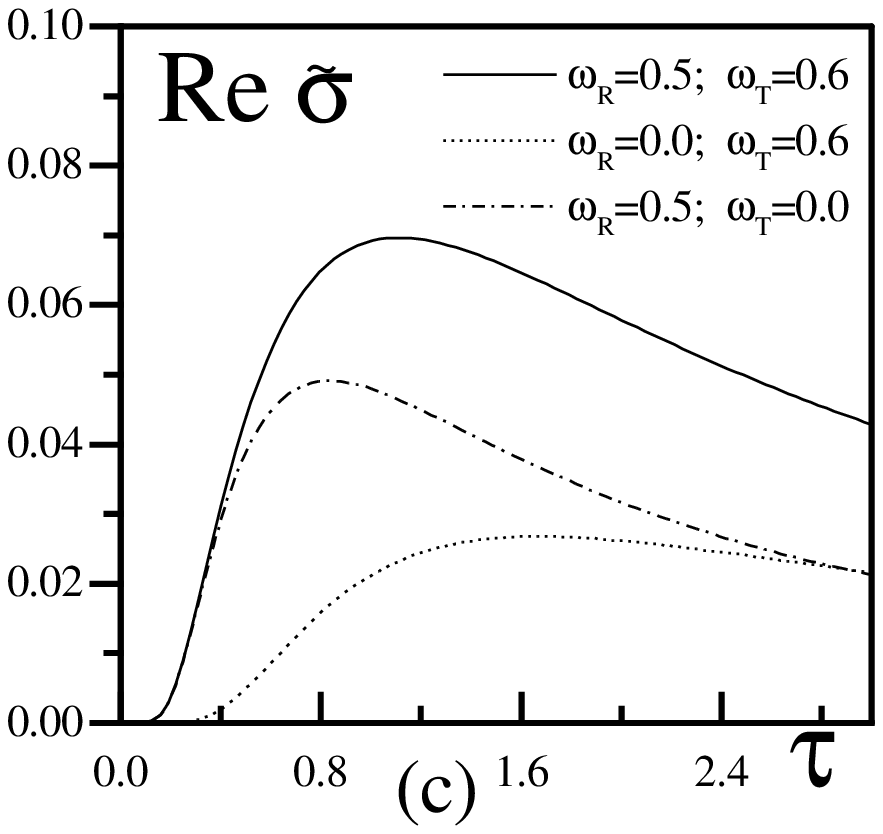}}
\caption{}
\label{fig9}
\end{figure}


\begin{references}

\bibitem{hadzi} D.~Had\v{z}i, J.~Mol.~Struct. {\bf 177}, 1 (1988).

\bibitem{kawada} A.~Kawada, A.~R.~McGhie and M.~M.~Labes, 
J.~Chem.~Phys. {\bf 52}, 3121 (1970).

\bibitem{onsager} M.-S.~Chen, L.~Onsager, J.~Bonner and J.~Nagle,
J.~Chem.~Phys. {\bf 60}, 405 (1974).

\bibitem{belushkin} A.~V.~Belushkin, C.~J.~Carlile and L.~A.~Shuvalov,
Ferroelectrics {\bf 167}, 21 (1995).

\bibitem{zetterstrom} P.~Zetterstr\"{o}m, A.~V.~Belushkin, 
R.~L.~McGreevy and L.~A.~Shuvalov, Solid State Ionics {\bf 167}, 21 (1999).

\bibitem{pietraszko} A.~Pietraszko, B.~Hilczer and A.~Pawlowski,
Solid State Ion. {\bf 119(1-4)}, 281 (1999)

\bibitem{pietraszko2} A.~Pietraszko, K.~\L ukaszewicz and M.A.~Augustyniak,
Acta Cryst. C {\bf 48}, 2069 (1992); K.~\L ukaszewicz, A.~Pietraszko
and M.A.~Augustyniak, {\it ibid.} {\bf 49}, 430 (1993);
A.~Pietraszko and K.~\L ukaszewicz, Bull. Polish Acad. Sci. 
{\bf 41}, 157 (1993).

\bibitem{pavlenko} N.~Pavlenko, J.Phys.:~Condens.~Matter {\bf 11}, 
5099 (1999).

\bibitem{kravchenko} S.~V.~Kravchenko, G.~V.~Kravchenko, J.~E.~Furneaux, 
V.~N.~Pudalov and M.~D'Iorio, Phys.~Rev.~B {\bf 50} 8039 (1994).


\bibitem{jps} I.~V.~Stasyuk, O.~L.~Ivankiv and N.~Pavlenko 
J.~Phys.~Stud. {\bf 1}, 418 (1997).

\bibitem{polarons} Yu.~A.~Firsov. {\it Polarons}, (Nauka, Moscow, 1975, in 
Russian).

\bibitem{salman} F.~E.~Salman, B.~Hilczer and Cz.~Pawlaczyk, Japanese 
J.~Appl.~Phys. {\bf 24}, 668 (1985).

\bibitem{padmanabhan} V.~M.~Padmanabhan and R.~Balasubramanian, 
Acta Cryst. {\bf 22}, 532 (1967).

\bibitem{kubo1} Kubo R 1957 {\it J. Phys. Soc. Jpn.} {\bf 12}, 570

\bibitem{reik} H.~G.~Reik and D.~Heese, J.Phys. Chem. Solids
{\bf 28} 581 (1967).


\end{references}
\end{document}